\documentclass[11pt,a4paper]{article}
\usepackage{moriond,epsfig}

\def\Journal#1#2#3#4{{#1} {\bf #2}, #3 (#4)}

\newcommand{\eV}{\,\mathrm{eV}}
\newcommand{\MeV}{\,\mathrm{MeV}}
\newcommand{\GeV}{\,\mathrm{GeV}}

\newcommand{\Mpc}{\,\mathrm{Mpc}}
\newcommand{\Gpc}{\,\mathrm{Gpc}}
\newcommand{\cm}{\,\mathrm{cm}}

\newcommand{\ud}{\mathrm{d}}
\newcommand{\tu}[1]{{\mathrm{#1}}}
\newcommand{\anti}[1]{\overline{#1}}

\begin{document}
\vspace*{4cm}
\title{Propagation of UHECRs}

\author{ Daniel De Marco }

\address{Bartol Research Institute, University of Delaware\\
Newark, DE 19716, U.S.A.}

\maketitle\abstracts{
In this general introduction to Ultra High Energy Cosmic Rays (UHECRs)
we discuss the propagation of UHE protons and the GZK feature that is
expected approaching $10^{20}\eV$ for homogeneously distributed sources.
We also briefly present the effects of the propagation on other
particles that can play the role of UHECRs. With the help of numerical
simulations for the propgation of UHECRs, we show that the GZK feature
cannot be accurately determined with the small sample of events with
energies $\sim10^{20}\eV$ detected thus far by the largest two
experiments, AGASA and HiRes.}

\section{Introduction}

In the past ninety years of cosmic rays research there has been a
constant search for the \emph{end of the cosmic-rays spectrum} and it
has long been thought that this end would be determined by the highest
energy that cosmic accelerators might be able to achieve. Despite the
continuous search, no end of the spectrum was found. In 1966, right
after the discovery of the cosmic microwave background (CMB), it was
understood~\cite{GZK} that high energy protons would interact
inelastically with the photons of the CMB and produce pions. For
homogeneously distributed sources this would cause a flux suppression,
called the \emph{GZK cutoff}: for the first time the end of the cosmic
ray spectrum was related to a physical process rather than to
speculations on the nature of the accelerators. Moreover, for the first
time, the end of the cosmic ray spectrum was predicted to be at a rather
well defined energy, around $10^{20}\eV$, where the so-called photo-pion
production starts to be kinematically allowed.

UHECRs can be of various nature and during their propagation over
cosmological distances they suffer different kinds of energy losses.
In this paper we consider most of the particles that could play the role
of UHECRs and we review the processes affecting their propagation. In \S
2 we discuss the propagation of protons, heavy nuclei, photons and
neutrinos. In \S 3 we show that the two largest experiments operating up
to now, AGASA and HiRes, have a too small statistic to provide a
conclusive answer about the presence or absence of the GZK feature in
the UHECRs spectrum. We conclude in \S 4.

\section{Propagation of UHECRs in the cosmic photon background}

\subsection{Protons}

There are three sources of energy loss for ultra high energy protons
propagating over cosmological distances: the expanding universe
redshift, pair production ($p\gamma\rightarrow p e^+e^-$) and photo-pion
production ($p\gamma\rightarrow\pi N$), each successively dominating as
the proton energy increases.

For protons the most important background is the CMB and the most
important process is the photo-pion production in which a nucleon of
sufficiently high energy sees, in its reference frame, the photons of
the CMB blue-shifted to $\gamma$-rays above the threshold energy for
photo-pion production,
$E^\tu{lab,thr}_\gamma=m_\pi+m_\pi^2/(2m_N)\simeq160\MeV$. The
cross-section for this process has a pronounced resonance just above
threshold, corresponding to the production of an intermediate state
$\Delta^+$ that immediately decays into a nucleon and a pion, whereas in
the limit of high energies it increases logarithmically with
$s=m_N^2+2m_NE_\gamma^\tu{lab}$, giving rise to multiple pion
production. For a background photon of energy $\epsilon$ in the cosmic
rest frame, defined as the frame in which the CMB is isotropic,
the threshold energy $E^\tu{lab,thr}_\gamma$ translates into a
corresponding threshold for the nucleon energy:
\begin{equation}
E_\tu{thr}=\frac{m_\pi}{(1-\cos\theta)\epsilon}\left(m_N+\frac{m_\pi}{2}\right) \simeq
6.8\!\cdot\!10^{19} \left(\frac{10^{-3}\eV}{\epsilon} \right)
\left(\frac{2}{1-\cos\theta}\right)
\eV\,.
\end{equation}
Typical CMB photon energies are of the order of $10^{-3}\eV$, giving a
threshold value of a few tens of EeV (for a head-on collision).

\begin{figure}
  \centering
  \begin{minipage}[t]{0.3\textwidth}
  \centering
  \includegraphics[width=\textwidth]{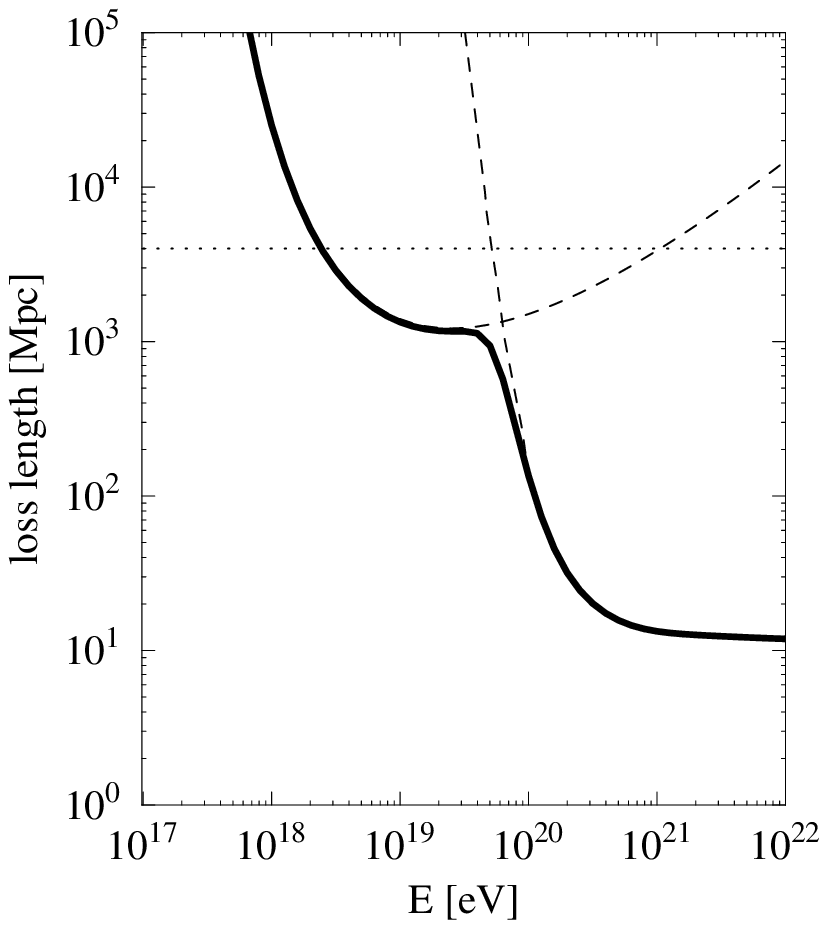}
  \caption{Solid line: loss length for photo-pion and photo-pair
  production for protons~\protect\cite{berez,stecker}.
  The dashed lines report the separate contributions of the two
  processes. The dotted line shows the redshift losses.
  }\label{fig:ll}
  \end{minipage}\qquad
  \begin{minipage}[t]{0.6\textwidth}
  \centering
  \includegraphics[width=\textwidth]{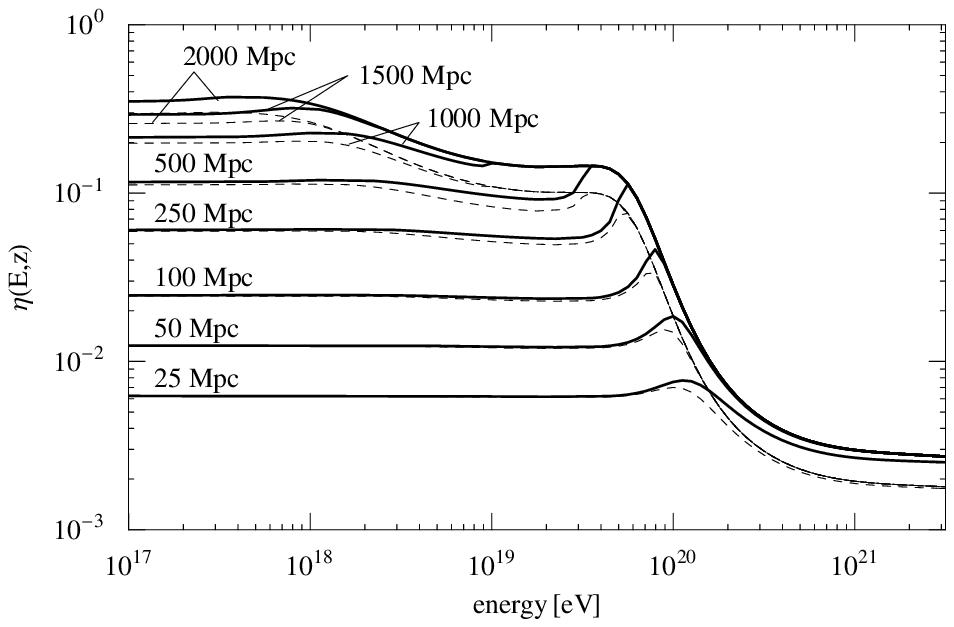}
  \caption{Modification factors as a function of the energy for
  many-source spectrum with $\gamma=2.1$ (solid lines) and $\gamma=2.7$
  (dashed lines). The sources are uniformly distributed up to the
  indicated distances. After Ref.~\protect\cite{berez}.
  }\label{fig:diffuse}
  \end{minipage}
\end{figure}

The interplay of this threshold with the Planck spectrum of the CMB
photons produces a very steep, exponential, curve for the interaction
length. The combination with the large inelasticity of the photo-pion
interaction (the mean inelasticity goes from $\sim0.13$ at threshold to
$\sim0.5$ at high energy, with large fluctuations) creates a very
efficient and rapid mechanism to reduce the nucleon energy and makes the
universe opaque to nucleons with energy above $\sim\!10^{20}\eV$ on
scales above $\sim\!100\Mpc$. The so-called GZK cutoff is due exactly to
this: the flux at earth of nucleons with energy below threshold, say
$5\!\cdot\!10^{19}\eV$, is due to contributions from (almost) all the
universe, from Fig.~\ref{fig:ll} the loss length at this energy is of
the order of $1\Gpc$, while doubling the energy the loss length is
reduced to $100\Mpc$, and only a small portion of the universe
contributes to the flux. Thus this change by a factor two in the energy
changes the loss length by almost an order of magnitude, which
translates in about the same ratio between the flux below
$5\!\cdot\!10^{19}\eV$ and above $10^{20}\eV$ if the sources have no
luminosity evolution and no local overdensity and there is no magnetic
field.

Below $\sim6\!\cdot\!10^{19}\eV$ the dominant loss mechanism for
protons becomes the production of electron-positrons pairs on the CMB,
$p\gamma\rightarrow p e^+e^-$, down to the corresponding threshold:
\begin{equation}
E_\tu{thr}=\frac{m_e}{\epsilon}\left(m_N+m_e\right)\simeq
4.8\!\cdot\!10^{17}\left(\frac{10^{-3}\eV}{\epsilon} \right)\eV\,.
\end{equation}
The interaction length for this process is much shorter than the one for
pion production, but on the other hand the inelasticity is much lower,
$\sim10^{-3}$. This makes the pair production loss length of the order
of Gpc (see Fig.~\ref{fig:ll}). The low inelasticity of pair production
allows in calculations to treat this process as a continuous energy
loss, whereas the pion production has to be treated as a discrete
process due to its large inelasticity.

The last important mechanism which dominates near and below the pair
production threshold is redshifting due to the expansion of the
universe. Fig.~\ref{fig:ll} shows the loss lengths for pion and pair
production as calculated in Ref.~\cite{berez}.

It is worth stressing that what has been named the GZK cutoff is in fact
a \emph{feature}~\cite{bbo} as the shape of the energy spectrum around
$10^{20}\eV$ depends on many unknowns. The modifications of the spectrum
shape due to the above-mentioned loss processes was first investigated
by Berezinsky and Grigorieva in Ref.~\cite{berez}. They calculated the
modification factor (basically the observed spectrum divided by the
injection spectrum) for a uniform distribution of sources up to a
maximum distance $d_\tu{max}$. Fig.~\ref{fig:diffuse} shows their
results for sources without cosmological evolution, $m\!=\!0$, for some
values of the maximum distance of the sources. For large $d_\tu{max}$,
which is the case we are interested in, the spectrum shows a steepening
followed by a flattening and then by a suppression. The flattening is
due to the interplay between the features produced by the pair and pion
production processes and it is an important feature for these spectra
since it has a characteristic shape. There are claims that this feature
has been observed in the experimental data~\cite{berez}, although it is
not yet clear if the feature in the data is due to this effect or if it
is due to the transition between the galactic and extra-galactic
components.

It is important to stress what we said above: what is generically
called GZK-cutoff is actually a \emph{feature} as the spectrum does not
end at $10^{20}\eV$ (see Fig.~\ref{fig:diffuse}), but has a flux
suppression that depends on many details such as the injection spectrum
of cosmic rays, the luminosity evolution of the sources, the local
overdensity of sources and the magnetic field strength in the
intergalactic medium. As an example, including the luminosity evolution
makes the sources at high redshift brighter that the nearby ones and
this enhances the flux suppression, while a local overdensity of sources
has the opposite effect~\cite{bbo}; a flatter spectrum produces a lesser
attenuation than a steeper one and the strength of the magnetic field in
the intergalactic medium con produce many interesting features, see for
example Ref.~\cite{stanev}.

\subsection{Heavy Nuclei}

For nuclei the situation is slightly different: the dominant loss
process above about $10^{19}\eV$ is photodisintegration in the CMB and
IR background (IRB) due to the giant dipole resonance, followed at lower
energy by the pair production. The photo-pion production process is
negligible, except for light nuclei at very high
energies~\cite{nuclei,roulet}. Indeed, for a nucleus of mass number $A$
and charge $Z$, the energy loss length for pion production is roughly
the same one of a nucleon with identical Lorentz factor. This is due to
the fact that the cross section for pion production is approximately
proportional to the mass number $A$, while the inelasticity is
proportional to $1/A$. For pair production we got a different behavior
because, while the inelasticity is proportional to $1/A$ as before, the
cross section is proportional to $Z^2$ resulting in an energy loss
length lower by a factor $A/Z^2$ with respect to a proton with the same
Lorentz factor. Since $Z\!\sim\!A/2$, the ratio of the photo-pair and
photo-pion production increases roughly linearly with $Z$~\cite{czs}.

\begin{figure}
\centering
\includegraphics[width=0.7\textwidth]{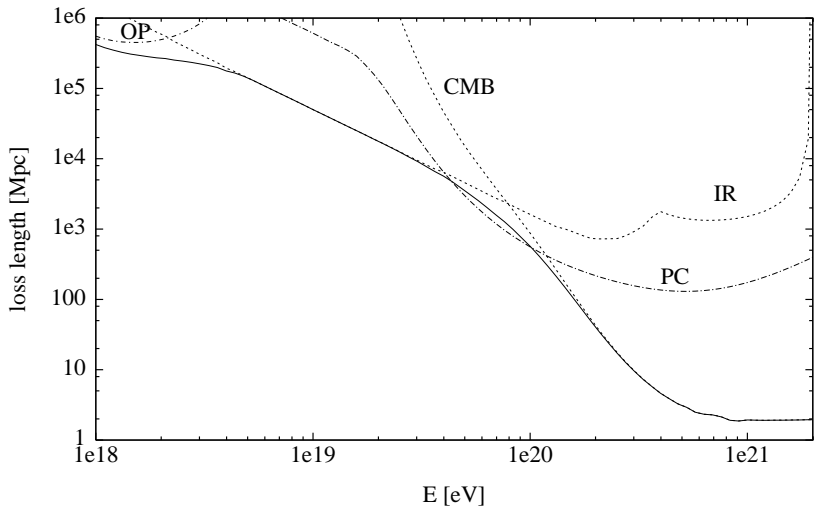}
\caption{Effective energy loss length for Fe photodisintegration off
microwave (CMB), infrared (IR) and optical (OP) photons, as well as the
total one (solid line) and the pair production loss length (PC). From
Ref.~\protect\cite{roulet}.}\label{fig:lossnuclei}
\end{figure}

The cross sections for photodisintegration $\sigma_{A,i}(\epsilon')$
contains essentially two regimes depending on $\epsilon'$, the photon
energy in the nucleus rest frame. At $\epsilon'<30\MeV$ there is the
domain of the giant dipole resonance and the disintegration proceeds
mainly by the emission of one or two nucleons. At higher energies, the
cross section is dominated by multi-nucleon emission for heavy nuclei
and is approximately flat up to $\epsilon'\!\sim\!150\MeV$. A useful
quantity to estimate the energy loss rate by photodisintegration is
given by the effective rate:
\begin{equation}
R^\tu{eff}_A=\frac{\ud A}{\ud t}=\sum_i i R_{A,i}\,.
\end{equation}
For photodisintegration, the average fractional energy loss results
equal to the fractional loss in mass number of the nucleus, $E^{-1}\ud
E/\ud t=A^{-1}\ud A/\ud t$, because the nucleon emission is isotropic in
the rest frame of the nucleus. Therefore during the photodisintegration
process the Lorentz factor of the nucleus is conserved, unlike the cases
of pair and pion production which involve the creation of new particles
that carry away energy. The energy loss time for photodisintegration is
then $A/R^\tu{eff}_A$. Fig.~\ref{fig:lossnuclei} shows separately the
different contributions to this quantity from CMB, IR and optical
photons for Fe nuclei, together with the total one (solid line) and the
pair creation loss length.

It is apparent that the optical background has no relevant effect, that
the IR one dominates the photodisintegration processes below
$10^{20}\eV$ and the CMB dominates above $10^{20}\eV$. The pair creation
rate is relevant for Fe energies
$4\!\cdot\!10^{19}\eV\div2\!\cdot\!10^{20}\eV$ ($\gamma$ factors
$\sim\!(1\div4)\cdot10^9$), for which the typical CMB photon energy in the
rest frame of the nucleus is above threshold ($>\!1\MeV$) but still well
below the peak of the giant resonance ($\sim\!10\div20\MeV$). The effect
of pair creation losses is to reduce the $\gamma$ factor of the nucleus,
obviously leaving $A$ unchanged~\cite{roulet}.

\begin{figure}
  \centering
  \begin{minipage}[t]{0.55\textwidth}
  \centering
  \includegraphics[width=\textwidth]{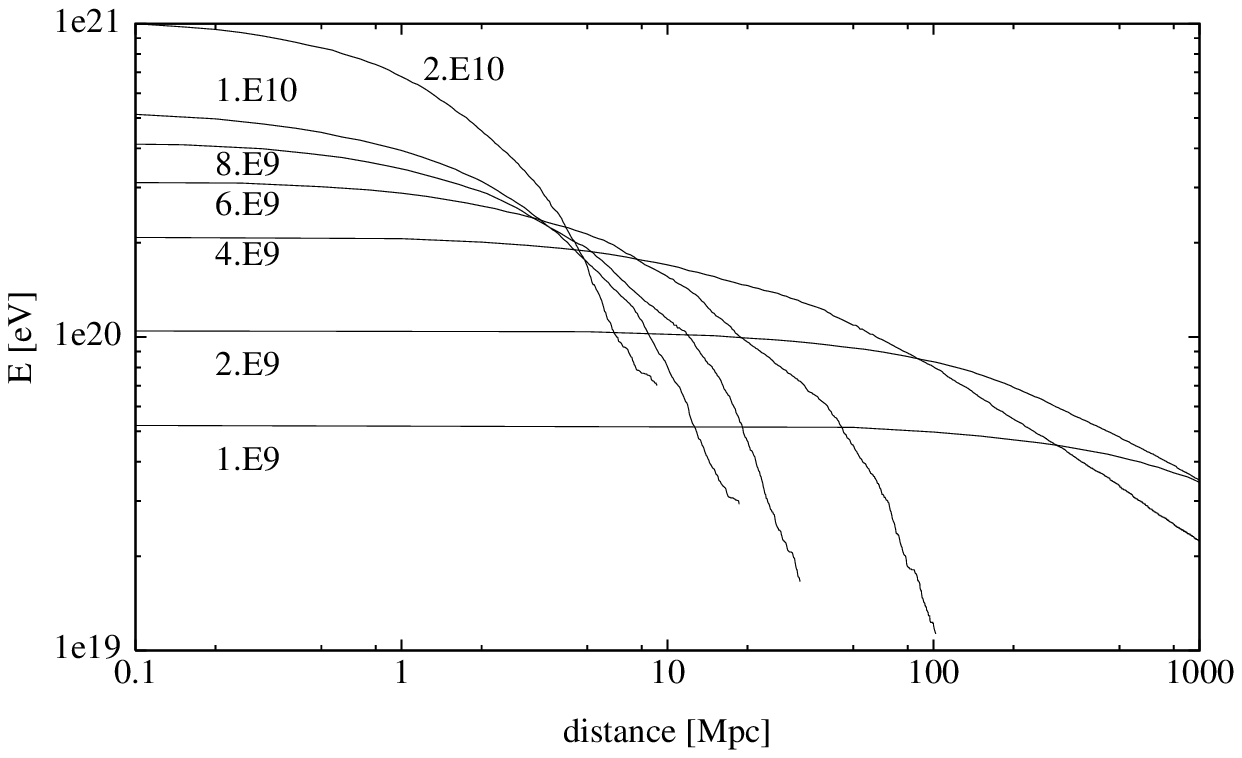}
  \caption{Average energy as a function of the propagation distance for
  particles that started as Iron nuclei with the indicated Lorentz
  factors. From Ref.~\protect\cite{roulet}.}\label{fig:evolnuclei}
  \end{minipage}\qquad
  \begin{minipage}[t]{0.4\textwidth}
  \centering
  \includegraphics[width=\textwidth]{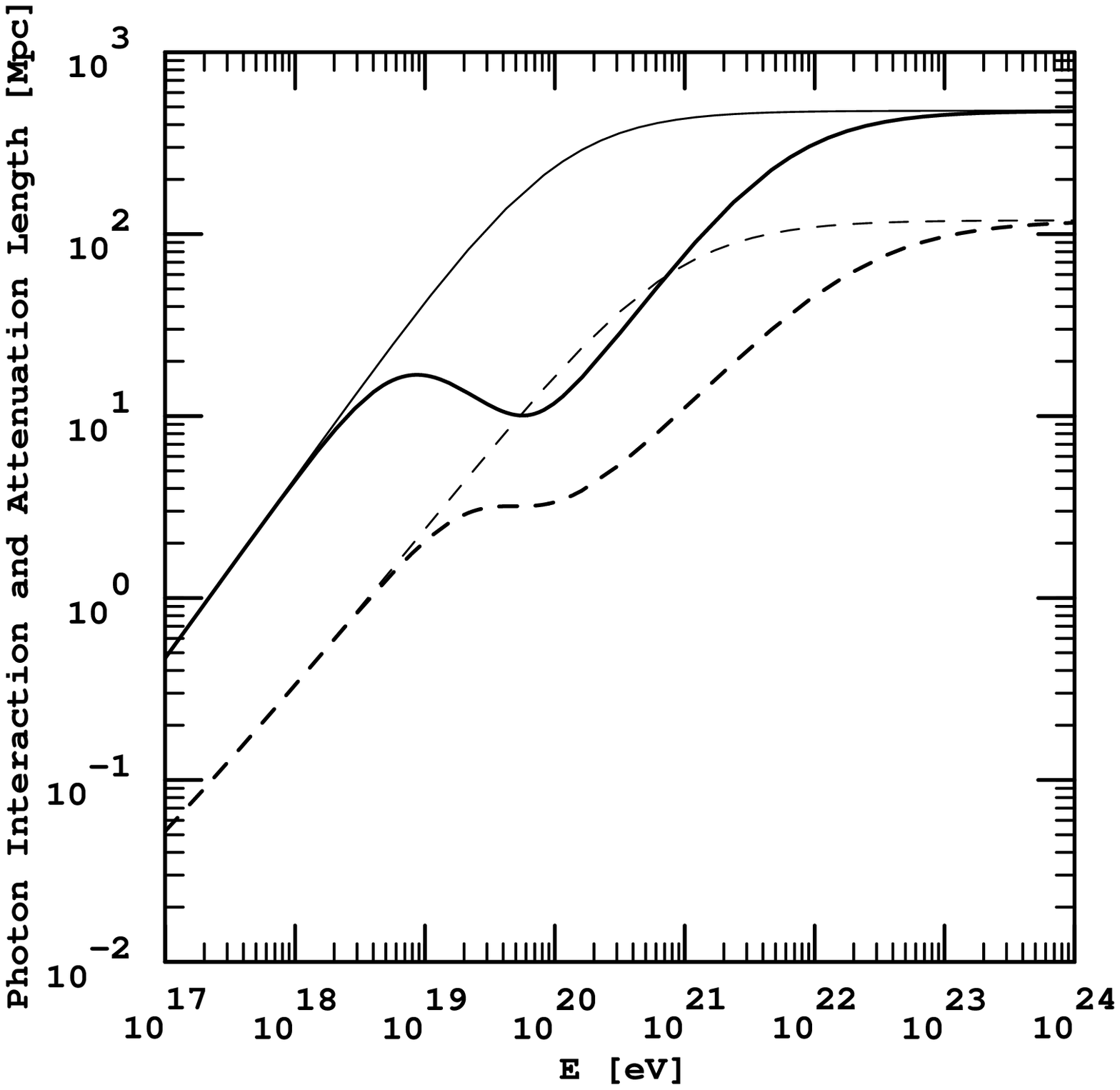} 
  \caption{Interaction lengths (dashed lines) and energy attenuation
  lengths (solid lines) of $\gamma$-rays in the CMB (thin lines) and in
  the combined CMB and URB (thick lines). The interactions taken into
  account are single and double pair production. From
  Ref.~\protect\cite{bs}.}\label{fig:gammalen}
  \end{minipage}
\end{figure}

We should not get fooled by the loss lengths in
Fig.~\ref{fig:lossnuclei} into thinking that if the loss length for a Fe
nucleus of $10^{20}\eV$ is $500\Mpc$, then we can receive on Earth a Fe
nucleus that started many hundred Mpc away. This is because the
corresponding interaction length is more than an order of magnitude
shorter and after every interaction the nucleus becomes lighter and
lighter and along with this the loss length for photodisintegration
becomes shorter and shorter. The net result~\cite{roulet}, as can be seen
in Fig.~\ref{fig:evolnuclei}, is that after $10\Mpc$ all the energies
are below $2\!\cdot\!10^{20}\eV$ and after $100\Mpc$ they are below
$10^{20}\eV$.

\subsection{Photons}

As in the case of UHE nucleons and nuclei, the propagation of UHE
photons (and electrons/pos\-i\-trons) is also governed by their interaction
with the cosmic photon background. The dominant interaction processes in
this case are the attenuation of UHE photons due to pair production (PP)
on the background photons ($\gamma \gamma_b \rightarrow e^+ e^-$), and
inverse Compton scattering (ICS) of the electrons (positrons) on the
background photons.

The $\gamma$-ray threshold energy for PP on a background photon of
energy $\epsilon$ is 
\begin{equation}
E_\tu{thr}=\frac{m_e^2}{\epsilon}\simeq2.6\!\cdot\!10^{11}\left(\frac{\epsilon}{\tu{eV}}\right)^{-1}\eV\,,
\end{equation}
whereas ICS has no threshold. In the high-energy limit, the total cross
sections for PP and ICS are: 
\begin{equation}
\sigma_\tu{PP}\simeq2\sigma_\tu{ICS}\simeq\frac{3}{2}\sigma_\tu{T}(m_e^2/s)\ln(s/2m_e^2)
\qquad (s\gg m_e^2)\,.
\end{equation}
For $s\ll m_e^2$, $\sigma_\tu{ICS}$ approaches the Thomson cross section
$\sigma_\tu{T}=8\pi\alpha^2/3m_e^2$ ($\alpha$ is the fine structure
constant), whereas $\sigma_\tu{PP}$ peaks near the threshold. Therefore,
the most efficient targets for electrons and $\gamma$-rays of energy $E$
are background photons of energy $\epsilon\simeq m_e^2/E$. For UHE this
corresponds to $\epsilon\sim<10^{-6}\eV\simeq100\,\tu{MHz}$. Thus,
radio background photons play an important role in UHE $\gamma$-ray
propagation through extragalactic space~\cite{bs}. Unfortunately, the universal
radio background (URB) is not very well known, mostly because it is
difficult to disentangle the Galactic and extragalactic components.

In the extreme Klein-Nishina limit, $s\gg m_e^2$, either the electron or
the positron produced in the process $\gamma\gamma_b\rightarrow e^+e^-$
carries most of the energy of the initial UHE photon. This leading
electron can then undergo ICS whose inelasticity (relative to the
electron) is close to 1 in the Klein-Nishina limit. As a consequence,
the upscattered photon which is now the leading particle after this
two-step cycle still carries most of the energy of the original
$\gamma$-ray, and can initiate a fresh cycle of PP and ICS interactions.
This leads to the development of an \emph{electromagnetic (EM) cascade}
which plays an important role in the resulting observable $\gamma$-ray
spectra. An important consequence of the EM cascade development is that
the effective penetration depth of the EM cascade, which can be
characterized by the energy attenuation length of the leading particle
(photon or electron/positron), is considerably greater than just the
interaction length (see Fig.~\ref{fig:gammalen})~\cite{bs}.

EM cascades play an important role particularly in some exotic models of
UHECR origin such as collapse or annihilation of topological defects in
which the UHECR injection spectrum is predicted to be dominated by
$\gamma$-rays. But, even if only UHE nucleons and nuclei are produced in
the first place, for example via conventional shock acceleration, EM
cascades can be produced by the secondaries coming from the decay of
pions which are created in interactions of UHE nucleons with the low
energy photon background~\cite{gamma}.

Most of the energy of fully developed EM cascades ends up below
$\sim\!100\GeV$ where it is constrained by measurements of the diffuse
$\gamma$-ray flux. Flux predictions involving EM cascades are therefore
an important source of constraints of UHE energy injection on
cosmological scales.

It should be mentioned that the development of EM cascades depends
sensitively on the strength of the extragalactic magnetic fields (EGMFs)
which is rather uncertain. The EGMF typically inhibits the cascade
development because of the synchrotron cooling of the $e^+e^-$ pairs
produced in the PP process. The energy lost through synchrotron
radiation does not, however, disappear; rather, it reappears at lower
energies and can even initiate fresh EM cascades.

\subsection{Neutrinos}

The propagation of UHE neutrinos is governed mainly by their
interactions with the relic neutrino background (RNB). The interaction
energies are typically smaller than electroweak energies even for UHE
neutrinos and then the cross sections are given by the Standard Model of
electroweak interactions which are well confirmed experimentally.
Physics beyond the Standard Model is not expected to play a significant
role in UHE neutrino interactions with the low-energy relic backgrounds.
Despite the neutrino-neutrino cross section are at least a few order of
magnitude smaller than the neutrino-nucleon ones, the latter
interactions are negligible compared to interactions with the RNB
because the RNB particle density, $\sim100\cm^{-3}$ per family, is about
10 orders of magnitude larger than the baryon density.

The $\nu\anti{\nu}$ annihilation mean free path is of the order of
$\lambda_\nu=(n_\nu\sigma_{\nu\anti{\nu}})^{-1}\simeq4\!\cdot\!10^{28}\cm$,
just above the present size of the horizon ($H_0^{-1}\sim10^{28}\cm$).
The neutrino is the only known stable particle that can propagate
through the universe essentially uninhibited even at the highest
energies. This has lead to the speculation that neutrinos could be
indeed the super-GZK primaries. However, in the Standard Model a
neutrino incident vertically in the atmosphere would pass through it
uninhibited, never initiating an extensive air shower. Consequently, for
these scenarios to work, one has to postulate new interactions so that
these neutrinos acquire a strong cross section above $10^{20}\eV$.

An interesting situation arises if the RNB consists of massive neutrinos
with $m_\nu\!\simeq\!1\eV$: such neutrinos would constitute hot dark
matter which is expected to cluster, for example, in galaxy clusters.
This would potentially increase the interaction probability for any
neutrino of energy within the width of the $Z^0$ resonance at
$E=M_Z^2/2m_\nu=4\!\cdot\!10^{21}(\tu{eV}/m_\nu)\eV$. 
It has been suggested that the stable end products
of the \emph{Z-bursts} induced at close-by distances
($\sim<\!50\Mpc$) from Earth may explain the highest energy cosmic
rays~\cite{zburst}. The problem with these proposals is however
that they require a very high flux of UHE neutrinos to begin with and
this makes Z-burst above GZK energies more likely to play a role in the
context of non-accelerating scenarios. For further information see
Ref.~\cite{bs} and references therein.

It is important to point out that the only conventional/assured source
of UHE neutrinos is the GZK effect itself. The neutrinos are the result
of the decay of the pions produced in the $p\gamma$ interaction. The
flux however is not very high and the detection is quite difficult. For
further informations see
Refs.~\cite{bs,neutrinos}.

\section{AGASA and HiRes: is there a discrepancy?}

AGASA and HiRes are, up to now, the two experiments with the larger
exposure for the detection of UHECRs. They reported however apparently
conflicting results. The two reported spectra appear: 1) to have a
systematic offset at low energy and 2) to differ above $10^{20}\eV$
where AGASA shown no hint of the GZK-suppression whereas HiRes seems to
be consistent with it. It has been shown~\cite{gzk} that a systematic
overestimate of the AGASA energies by 15\% and a corresponding
underestimate of the HiRes energies by the same amount would in fact
bring the two data sets in a much better agreement in the region below
$10^{20}\eV$. In Ref.~\cite{gzkbis} we applied our Monte Carlo
simulation~\cite{gzk,gzkbis,multi} to investigate the discrepancies at
high energy and we found that:
\begin{itemize}
  \item assuming a uniform distribution of sources, the AGASA spectrum
    is reproduced, in a conventional scenario where the average spectrum
    has the GZK suppression, with a probability of
    $\sim6\!\cdot\!10^{-4}$ ($\sim3\sigma$).
  \item assuming the presence of the 15\% systematic error, the shifted
    AGASA data are reproduced with a probability of
    $\sim6\!\cdot\!10^{-3}\div10^{-2}$ ($2.5\sigma\div2\sigma$).
  \item the HiRes data are reproduced, in a scenario without a GZK
    suppression\footnote{To mimic a scenario without GZK suppression we
    used the AGASA dataset as template.}, with a probability of
    $\sim2\%$, ($\sim2\sigma$).
\end{itemize}
It is important to stress that in order to properly do these
calculations one has to take into account the statistical error in the
energy determination. Due to the steeply falling spectrum and to the
expected change of slope around $10^{20}\eV$ the statistical error in
the energy determination changes the average expected number of events
above $10^{20}\eV$ by $\sim1$ and with the present limited statistics
even a difference of one event is important~\cite{gzkbis}.

\begin{figure}
  \centering
  \includegraphics[width=0.9\textwidth]{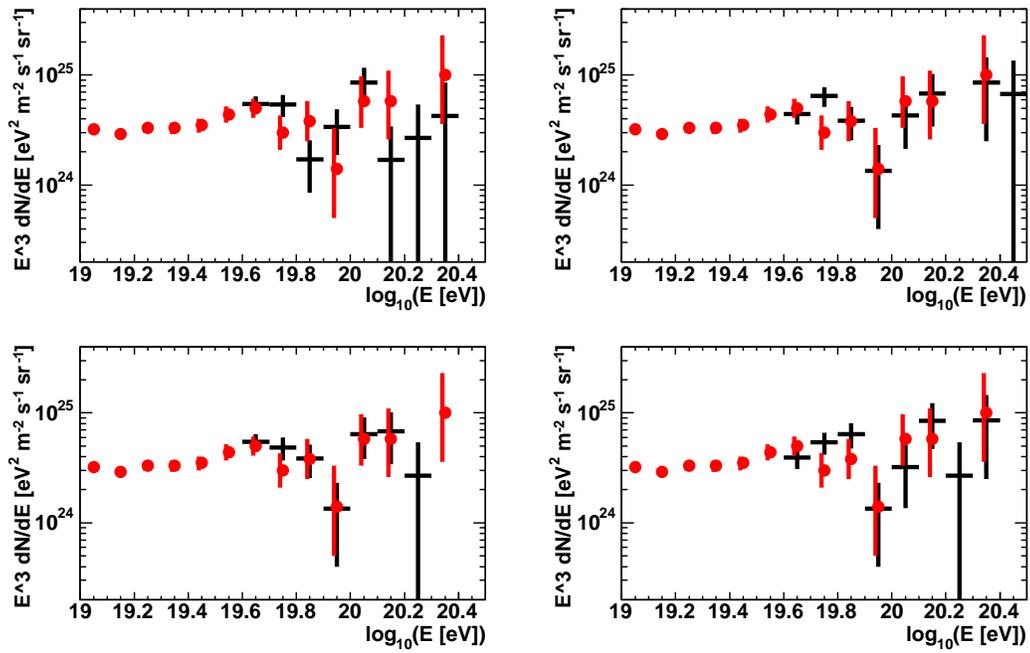}
  \vspace*{12pt}
  \caption{In the above panels we plot 4 of the 18 simulations that have
  11 or more events above $10^{20}\eV$. The black crosses are the
  simulation results. The red dots with errorbars are the AGASA data
  superimposed for comparison. The errorbars in the AGASA data are
  slightly shifted left to avoid covering up the black
  crosses.}\label{fig:spectra} 
\end{figure}

In Fig.~\ref{fig:spectra} we plot the spectra of some of the simulated
AGASA realizations that produced 11 or more events above $10^{20}\eV$.
It is striking the resemblance of the simulated spectra to the AGASA
one: all of them show no evidence of the GZK suppression. This shows
that the AGASA spectrum is far from being impossible, even if the
average cosmic ray spectrum can be expected to show a GZK feature.

From the above points we can conclude that neither AGASA nor HiRes have
enough statistical power to prove the presence or absence of the GZK
feature in the spectrum of UHECRs. A new generation of experiments is
needed to finally provide a conclusive answer to this question.

\section{Conclusions}

We considered several particles that can possibly play the role of
UHECRs and we discussed the energy loss processes that affect their
propagation over cosmological distances. We showed that for all
considered particles the energy losses above $\sim10^{20}\eV$ become
so severe that they cannot propagate over distances larger than about
$100\Mpc$ without reducing their energy below $10^{20}\eV$.
The effect of these energy losses on the spectrum of UHECRs is the so
called GZK-suppression, due to the fact that below $10^{20}\eV$ almost
all the universe is contributing to the observed flux whereas above
$10^{20}\eV$ we receive contributions only from sources not too far
away, $\sim100\Mpc$. We stressed the point that this is not a cutoff,
but a {\it feature} since the spectrum does not end at $10^{20}\eV$, but
it is only suppressed, and the amount of this suppression depends on
many unknowns such as: the luminosity evolution of the sources, their
local overdensity and the magnetic field strength in the intergalactic
medium. 

We showed that the present sets of data are not enough to determine
whether this GZK suppression is present or not in the observed spectrum
and that we need a new generation of experiments to have a conclusive
answer to this question.

\section*{Acknowledgments}
This research is funded in part by NASA grant NAG5-10919.

\end{document}